%% file: gene5.tex
\documentclass[12pt]{article}

\input macros

\begin{document}
\begin{center}
{\Large {\bf Logic Programming with Macro Connectives}}
\\[20pt] 
{\bf Keehang Kwon}\\
Dept. of Computer  Engineering, DongA University \\
Busan 604-714, Korea\\
  khkwon@dau.ac.kr\\
\end{center}

\noindent {\bf Abstract}: 
Logic programming such as Prolog is often sequential and slow because each execution step processes only a 
single, $micro$ connective. To fix this problem, we propose to use $macro$ connectives
as the means of improving both readability and performance. 


{\bf keywords:} Prolog, macro connectives, synthetic connectives.

\newcommand{\muprolog}{{Prolog$^{macro}$}}
\newcommand{\somes}{\some \tilde{x}}
\newcommand{\somep}{\some \ddot{x}}
\newcommand{\alls}{\all \tilde{x}}
\newcommand{\allp}{\all \ddot{x}}
\renewcommand{\intp}{ex} 

\section{Introduction}\label{sec:intro}
Modern imperative languages such as Java, Perl  support macro connectives to
improve readability and performance of a program. The $switch$ statement is such an
example. To be precise, the $switch$  statement ($n$-ary branch) is $redundant$ in the sense that it can be converted to 
the  $if$-$then$-$else$ (binary branch) statement. However,  this $switch$  statement 
has proven essential in many programs.

Unfortunately, the situation is quite different in logic and logic programmming.
For example, first-order logic (FOL)  requires logical connectives $\land,\lor$ to be restricted to 
binary.
 For example, $\land(A,B,C)$ ($(A \land B\land C)$ in infix notation)
 must be written as either $\land(A,\land(B,C))$ or $\land(\land(A,B),C)$.
Similarly, it requires  $\vec{x}$ in $\all \vec{x}, \exists \vec{x}$ to be  a single
variable where $\vec{x} = x_1,\ldots,x_n$. 
This restriction is rather unnatural and has some unpleasant consequences, known
as {\it syntactic bureaucracy}.

\begin{itemize}

\item It  increases the complexity of formulas and, therefore,
makes formulas more difficult to read and write.

\item  It  makes proof search   more sequential and less parallel.
 It  forces proof steps that are parallel in nature to
 be written in a sequential order.

\item  It makes (already highly nondeterministic) proof search  less atomic and more
 nondeterministic.

\end{itemize}

\newcommand{\folp}{FOL$^+$}
\newcommand{\lkp}{LK$^+$}

To fix this problem of syntactic bureaucracy, we extend FOL to \folp\ to include the following
macro formulas (called generalized conjunction/disjunction, block universal/existential
 quantifiers, respectively):

\begin{itemize}

\item $\land(F_1,\ldots,F_n)$, $\lor(F_1,\ldots,F_n)$ are formulas for $i = 2,3,\ldots$.

\item $\exists\vec{x} F$, $\exists\bar{x} F$, $\somes F$, $\somep F$,
  $\all\vec{x} F$, $\all\bar{x} F$, $\alls F$, $\allp F$  are formulas if $F$ is a
  formula. 
\end{itemize}
\noindent In the above (and in the sequel as well), $\vec{x}$ represents
$x_1,\ldots,x_n$, $\bar{x}$ represent $\lb x_1,\ldots,x_n \rb$,
$\tilde{x}$ represents $[x_1,\ldots,x_n]$ and
  $\ddot{x}$ represents $\{ x_1,\ldots,x_n \}$. 
The meaning of these formulas is  defined by the following. Note that it is
based on the game semantics \cite{Jap03}, an extension to
the traditional true/false semantics.

\begin{itemize}

\item  $\exists \vec{x} F$ is identical to $\exists x_1 \ldots \some x_n F$. That is,
      each $x_i$ must be processed in that order.  This connective can be seen
  as a concise version of the latter.  This formula is called a  sequential
  existential
  quantifier.

\item  $\exists \bar{x} F$ is a new connective and is
      identical to $\exists \vec{x}  F$, except that each $x_i$ can be
        processed in any order. This formula is called a parallel
  existential
  quantifier.
    
\item $\somes F$ is identical to $\exists x_1 \ldots \some x_n F$ with
  the additional constraint that, in the former,
  $x_1,\ldots,x_n$ must be processed
  {\it consecutively} in that order. $\somes F$ is called a block sequential
  existential
  quantifier.
  
\item $\somep F$ is identical to $\somes  F$ with
  the difference that, in the former $x_1,\ldots,x_n$ must be processed
  {\it consecutively} but in arbitrary order. $\somes F$ is called a block
  parallel
  existential
  quantifier.

\end{itemize}
\noindent $\all \vec{x} F$, $\all \bar{x} F$, $\alls F$, $\allp F$ are similarly defined.

As can be seen above, macro connectives often lead to new connectives which have no counterpart
in micro connectives. Some new properties -- consecutiveness, order independence, etc --
often emerges when we deal with macro connectives.
These new connectives are introduced to deal with such emergent properties.

These new connectives provide a useful tool for capturing real-life interactive systems
such as airline reservation systems. These systems typically require complex and diverse
forms of interaction with the user including order-dependent/independent interactions,
consecutive/nonconsecutive interactions.

A sequent calculus for \folp\  can be easily obtained by extending the standard
sequent rules of Gentzen's LK for $\land,\lor,\all, \some$ with new synthetic rules.
Thus, in the new calculus, a small consecutive local inference steps can be combined into a
 single 
synthetic step, thus making proof search more parallel and more deterministic.

In this paper, our focus is on applying this idea to logic programming for improved 
conciseness and improved performance. 

For example, we adopt the following operational semantics for $\land$ and $\lor$.

\begin{itemize}

\item $\intp(D, \land(G_1, \ldots,  G_n))$ {\rm if}\ 
       $\intp(D,G_1)\ pand\ \ldots pand\ \intp(D,G_n)$

\item $\intp(D, \lor(G_1, \ldots G_n))$ {\rm if}\  
$\intp(D,G_1)\ por\ \ldots por\ \intp(D,G_n)$
\end{itemize}
\noindent where $\lor$ represents classical disjunction,
 $pand$  represents a parallel conjunction\cite{Jap03}, and $por$  represents a 
parallel disjunction\cite{Jap03}.

    This paper proposes  \muprolog, an extension of Prolog with macro connectives.
The remainder of this paper is structured as follows. We describe \muprolog\
  in
the next section. 
Section~\ref{sec:conc} concludes the paper.

\section{The Language}\label{sec:logic}

The language is a version of Horn clauses
 with macro connectives. 
It is described
by $G$- and $D$-formulas given by the syntax rules below:
\begin{exmple}
  \>$G ::=$ \>   $A \sep  \land(G_1,\ldots,G_n)   \sep  \lor(G_1,\ldots,G_n)\sep
  \some \vec{x} G \sep   \some \bar{x} G\sep \somes G \sep \somep G$ \\   \\
  \>$D ::=$ \>  $A  \sep G \supset A\ \sep \all \vec{x}  D \sep \all \bar{x}  D \sep \alls   D \sep \allp   D \sep  \land(D_1,\ldots,D_n) $
\end{exmple}
\noindent
\newcommand{\sync}{up}
\newcommand{\async}{down}

In the rules above, 
$A$  represents an atomic formula.
A $D$-formula  is called a  Horn
 clause with macro connectives. 

The logic programming paradigm such as Prolog was originally founded on the resolution method.
But this approach was difficult to extend to  richer logics. The use of 
sequent calculus allows us to overcome this limit.
 In particular, uniform proofs \cite{MNPS91} 
allows us to execute logic programs in an efficient way by integrating two separate phases --
the proof phase and the execution
 phase -- into a single phase. We adopt this approach below.

\newcommand{\bc}{bc}

 Note that execution  alternates between 
two phases: the goal-reduction phase 
and the backchaining phase. 
In  the goal-reduction phase (denoted by $\intp(D,G)$), the machine tries to solve a goal $G$ from
a clause $D$  by simplifying $G$. 
If $G$ becomes an atom, the machine switches to the backchaining mode. 
In the backchaining mode (denoted by $bc(D_1,D,A)$), the machine tries 
to solve an atomic goal $A$ 
by first reducing a Horn clause $D_1$ to simpler forms  and then 
backchaining on the resulting 
 clause (via rule (1) and (2)). 

\begin{defn}\label{def:semantics}
Let $G$ be a goal and let $D$ be a program.
Then the notion of   executing $\lb D,G\rb$ -- $\intp(D,G)$ -- 
 is defined as follows:

\begin{numberedlist}

\item  $\bc(A,D,A)$. \% This is a success.

\item    $\bc((G_0\supset A),D,A)$ if 
 $\intp(D, G_0)$. \% backchaining

\item    $\bc(\land(D_1,\ldots,D_n),D,A)$ if   $\bc(D_1,D,A)$ por $\ldots$ por\ $\bc(D_n,D,A)$.
  
\item    $\bc(\all x_1,\ldots,x_n D_1,D,A)$ if   $\bc(\all x_2\ldots x_n  [t_1/x_1]D_1,D, A)$. Thus it processes only $x_1$.

\item    $\bc(\all \lb x_1,\ldots,x_n\rb D_1,D,A)$ if   $\bc(\all \lb x_1\ldots x_{i-1},x_{i+1},\ldots,
  x_n\rb [t_i/x_i]D_1,D, A)$. Thus it processes only $x_i$ for some $i$.  

\item    $\bc(\alls D_1,D,A)$ if   $\bc([t_1/x_1]\ldots [t_n/x_n]D_1,D, A)$ where  $t_1,\ldots,t_n$ are terms.  Thus, the variables $x_1,\ldots,x_n$ are processed both consecutively
  and sequentially.
  
\item    $\bc(\allp D_1,D,A)$ if   $\bc([t_1/x_1,\ldots,t_n/x_n]D_1,D, A)$ where  $t_1,\ldots,t_n$ are terms. 
  Thus, the variables $x_1,\ldots,x_n$ are processed both consecutively
  and in parallel.

\item    $\intp(D,A)$ if   $\bc(D,D, A)$. \% switch to backchaining mode


\item  $\intp(D, \land(G_1,\ldots,G_n))$  if $\intp(D,G_1)$  $pand$ $\ldots$ $pand$
  $\intp(D,G_n)$.

\item  $\intp(D, \lor(G_1,\ldots,G_n))$  if $\intp(D,G_1)$  $por$ $\ldots$ $por$
  $\intp(D,G_n)$.

\item $\intp(D,\exists\ x_1,\ldots,x_n   G)$  if $\intp(D,\exists\ x_2,\ldots,x_n [t_1/x_1]G)$
where   $t_1$ is a term. Thus, it processes only $x_1$.

\item $\intp(D,\exists\ \lb x_1,\ldots,x_n\rb   G)$  if $\intp(D,\exists\ \lb x_1,\ldots,x_{i-1},
  x_{i+1},\ldots,x_n\rb [t_i/x_i]G)$
  where   $t_i$ is a term. Thus, it processes only $x_i$ for some $i$.
  
\item $\intp(D,\somes   G)$  if $\intp(D,[t_1/x_1]\ldots [t_n/x_n]G)$
where   $t_1,\ldots,t_n$ are terms. Thus, the variables $x_1,\ldots,x_n$ are processed both consecutively
  and sequentially.

\item $\intp(D,\somep   G)$  if $\intp(D,[t_1/x_1,\ldots,t_n/x_n]G)$
where   $t_1,\ldots,t_n$ are terms. Thus, the variables $x_1,\ldots,x_n$ are processed both consecutively
  and in parallel.

\end{numberedlist}
\end{defn}

\noindent  
These rules are straightforward to read. Note that the
use of block  quantifiers makes it easy to substitute terms for
$x_1,\ldots,x_n$  by traversing formulas only once.

 As an example, consider the following  specification for computing binomial
   coefficients, denoted by $c(n,k,z)$.

\begin{exmple}
$\all N\ c(N,1,N).$ \% select one out of n \\  
$\all N\ c(N,N,1).$ \% select n out of n \\ 
$\all \{ N,K \}\ c(N,K,0)$ ${\rm :-}$ \> \hspace{8em} $N < K.$  \\ 
$\all \{ N,K,W,Z \}\ c(N,K,W+Z)$ ${\rm :-}$ \> \hspace{14em} $c(N-1,K-1,W) \land  c(N-1,K,Z).$
\end{exmple}
\noindent The above program is a little simpler and more efficient
than Prolog due to the use of
block universal quantifiers. The correctness of the above program
is guaranteed from the focalization property of traditional logic.

While it does not seem like much, it is easy to 
see that the benefits of using macro connectives will be substantial for highly complex 
formulas.

\section{Conclusion}\label{sec:conc}

In this paper, we have considered an extension to Prolog\cite{Bratko} with some 
macro connectives.  This extension makes Prolog programs easier to read, write and
execute.

Our macro connectives is a  simple yet practical subset of a wider class of  connectives
called {\em synthetic} connectives. These synthetic connectives -- proposed originally by 
Girard  -- is theoretically interesting and is based on the notion of focalization in 
linear logic. 
 In the near future, we plan 
to investigate the possibility of including these synthetic connectives into logic 
programming.

\bibliographystyle{plain}


\end{document}

%% file: macros.tex
\newenvironment{numberedlist}
{\begin{list}{\makebox[20pt]{\hss(\arabic{itemno})\enspace}}
             {\usecounter{itemno}\labelwidth 20pt}}{\end{list}}

\newcounter{itemno}

\newcounter{itemno1}

\newcounter{itemno2}

\newcounter{exno}

\newcounter{defno}

\newenvironment{defn}{\refstepcounter{defno}\medskip \noindent {\bf
Definition \thedefno.\ }}{\medskip}

\newcommand{\sep}{\;\vert\;}

\newcommand{\oprove}{\vdash\kern-.6em\lower.7ex\hbox{$\scriptstyle O$}\,}

\newcommand{\pderivation}{{\cal P}\kern -.1em\hbox{\rm -derivation}}
\newcommand{\pderivationl}{{\cal P}\kern -.1em\hbox{\em -derivation}}
\newcommand{\pderivable}{{\cal P}\kern -.1em\hbox{\rm -derivable}}
\newcommand{\pderivablel}{{\cal P}\kern -.1em\hbox{\em -derivable}}
\newcommand{\pderivations}{{\cal P}\kern -.1em\hbox{\rm -derivations}}
\newcommand{\pderivability}{{\cal P}\kern -.1em\hbox{\rm -derivability}}

\newcommand{\all}{\forall}
\newcommand{\some}{\exists}

\newsavebox{\lpartfig}
\newsavebox{\rpartfig}

\newenvironment{exmple}{
 \begingroup \begin{tabbing} \hspace{2em}\= \hspace{3em}\= \hspace{3em}\=
\hspace{3em}\= \hspace{3em}\= \hspace{3em}\= \kill}{
 \end{tabbing}\endgroup}

\newcommand{\lb}{\langle}
\newcommand{\rb}{\rangle}

\newcommand{\intp}{intp_o}
  
     {\\* \hspace*{\fill} \end{trivlist}}

%% file: gene5.bbl
\begin{thebibliography}{1}




\bibitem{Bratko}
I.~Bratko,   ``Prolog:programming for AI '',
 Addison Wesley, 2001 (3rd edition). 


\bibitem{Gir87}
J.Y.~Girard, ``Linear Logic'', Theoretical Computer Science, vol.50, pp.1--102, 
1987.

\bibitem{HM94}
J.~Hodas and D.~Miller, ``Logic Programming in a Fragment of Intuitionistic Linear Logic'', Information and Computation, vol.110, pp.327--365, 1994.


\bibitem{Jap03}
G.~Japaridze, ``Introduction to computability logic'', Annals  of Pure and
 Applied  Logic, vol.123, pp.1--99, 2003.



\bibitem{KK07}
E.~Komendantskaya and V.~Komendantsky, ``On uniform proof-theoretical operational semantics for logic programming'',  In J.-Y. Beziau and A.Costa-Leite, editors, Perspectives on Universal Logic, pages 379--394. Polimetrica Publisher, 2007.

\bibitem{Mil89jlp}
D.~Miller, ``A logical analysis of modules in logic programming'', Journal of
  Logic Programming, vol.6, pp.79--108, 1989.

\bibitem{MNPS91}
D.~Miller, G.~Nadathur, F.~Pfenning, and A.~Scedrov, ``Uniform proofs as a
  foundation for logic programming'', Annals of Pure and Applied Logic, vol.51,
  pp.125--157, 1991.

 








\end{thebibliography}
